\documentclass[twocolumn,pre,superscriptaddress]{revtex4-1}
\usepackage{graphicx}
\usepackage{amsmath}
\usepackage{dcolumn}
\usepackage{bm}
\usepackage{xcolor}
\usepackage{braket}
\usepackage{amssymb}
\begin{document}
\title{Non-equilibrium steady states of long-range coupled harmonic chains}
\author{Francesco Andreucci}

\affiliation{SISSA and INFN, Sezione di Trieste,
Via Bonomea 265, I-34136 Trieste,
Italy} 

\author{Stefano Lepri}
\affiliation{Consiglio Nazionale delle Ricerche, Istituto dei Sistemi Complessi, Via Madonna del Piano 10 I-50019 Sesto Fiorentino, Italy.} 
\affiliation{Istituto Nazionale di Fisica Nucleare, Sezione di Firenze, 
via G. Sansone 1, I-50019 Sesto Fiorentino, Italy}

\author{Stefano Ruffo}
\affiliation{SISSA and INFN, Sezione di Trieste,
Via Bonomea 265, I-34136 Trieste,
Italy} 
\affiliation{Consiglio Nazionale delle Ricerche, Istituto dei Sistemi Complessi, Via Madonna del Piano 10 I-50019 Sesto Fiorentino, Italy.}

\author{Andrea Trombettoni}
\affiliation{Department of Physics, University of Trieste, Strada Costiera 11, I-34151 Trieste, Italy}
\affiliation{SISSA and INFN, Sezione di Trieste,
Via Bonomea 265, I-34136 Trieste,
Italy} 
\affiliation{CNR-IOM DEMOCRITOS Simulation Center, Via Bonomea 265, I-34136 Trieste, Italy}

\begin{abstract}
%Non-equilibrium steady states of long-range interacting systems 
%are an interesting subject of investigation and received considerable attention in recent years. 
We perform a numerical study of transport properties 
of a one-dimensional  chain  
 with couplings  decaying as an inverse power 
$r^{-(1+\sigma)}$ of 
the inter-site distance $r$ and open boundary conditions, interacting
with tho heat reservoirs.
Despite its simplicity, the model displays highly nontrivial features 
in the strong long-range regime, $-1<\sigma<0$. 
At weak coupling with the reservoirs, the energy flux departs from the 
predictions of perturbative theory and displays anomalous superdiffusive
scaling of the heat current
with the chain size. We trace back this behavior to the transmission 
spectrum of the chain, which displays a self-similar structure 
with a characteristic sigma-dependent fractal dimension.
\end{abstract}

\maketitle 
\section{Introduction}
The main task of statistical mechanics is to relate the
microscopic interactions of a given system to its macroscopic 
%transport
properties. 
One typical instance is the context of heat transfer.
Suppose we apply a temperature gradient $\nabla T$ to a system, after a while the system will reach a stationary state characterized by the presence of a heat flux $\mathcal{J}$. The thermal conductivity $\kappa$ is defined in terms of these quantities as: 
\begin{equation} \label{fourier}
	\mathcal{J} = -\kappa \nabla T,
\end{equation}
In the case of diffusive transport, Fourier's law holds and $\kappa$ does not depend on the size of the system $N$ in the thermodynamic limit. This is typically the case for three-dimensional systems with short-range interactions.
We remark, however, that there is currently no generic way, given the microscopic properties of a system, to know whether Fourier's law holds or not.

A case in which Fourier's law is systematically violated is the case of harmonic interactions. For instance, for the harmonic crystal each phonon propagates freely and the transport is ballistic. This was showed for the first time for a chain with nearest-neighboors interactions in the seminal paper by Rieder, Lebowitz and Lieb \cite{RLL}. They found that the thermal conductivity $\kappa$ diverges as $\kappa \propto N$, $N$ being the number of particles in the chain. Moreover, the bulk temperature profile is flat, while Fourier's law would lead to a linear one. The non-equilibrium properties of quantum harmonic lattices 
have also been considered in the last decades \cite{Zuercher1990,
dhar2006,saito2000energy,asadian2013heat,freitas2014}.

Generally speaking, in harmonic lattices transport features are dictated by 
the spectral properties of both the thermal reservoirs and the system itself.
For instance in the case of disordered lattices displaying 
Anderson localization, the 
conductivity (or energy flux) depends on the localization
lengths, but also on the boundary conditions
\cite{visscher1971localization}, 
the spectral density
of the baths at low frequencies \cite{dhar2001heat} as well as
on the distribution and correlations of the random 
disorder \cite{herrera2015recovery,ash2020thermal}. 
For more general, non homogeneous harmonic networks, the spectral properties 
can be accounted by random matrix theory and can describe also
current fluctuations \cite{schmidt2013random}. 
This is even more striking for active (non-equilibrium) 
baths that can lead to non-trivial transport regimes 
even for the ordered harmonic chain \cite{santra2022activity}.
 
It %has
became progressively become clear that in one (and two) dimensions there are violations of Fourier's law also for nonlinear systems \cite{Lepri_review,DHARREV,Lepri2016,benenti2020,benenti2023non}, such as the Fermi-Pasta-Ulam-Tsingou (FPUT) chain. In one dimension, these violations manifest themselves as a 
power-law divergence of the thermal conductivity $\kappa$ with the system's size $\kappa \propto N^{\alpha}$. Transport in these cases is called anomalous. It is now clear that superdiffusive transport is a generic feature of non-linear one-(and two)-dimensional non-integrable systems conserving momentum, energy and stretch. %To conclude this very brief overview,
There are both numerical and analytical evidences that the exponent $\alpha$ can be used to identify different universality classes \cite{benenti2020}. For weakly non-integrable models the scenario may 
be more involved since quasi-particles may have very large mean-free paths \cite{dicintio2018transport,lepri2020tooclose}.

A further element of %complexity
interest is represented by the presence of forces that are not 
strictly local. Indeed, 
much less is known about systems with long-range interactions, that is, systems in which the inter-particle interaction scales with the particle distance $r$
as $V(r) \propto r^{-d-\sigma}$.
%This kind of interaction is far from being just an academic example. Indeed,
Several physical systems are characterized by long-range interaction, both classical (gravity, pure plasmas, $2d$ hydrodynamics) and quantum (dipolar systems and trapped atoms).
As a concrete experimental instance we mention 
trapped ion chains, where 
ions are confined in periodic arrays and interact with external
reservoirs  \cite{bermudez2013controlling,ramm2014energy}.
On a macroscale, effective long-range forces arise for tailored macroscopic systems like chain of coupled magnets \cite{moleron2019nonlinear} and the effects
of fluctuations and nonlinearity may be relevant.

Long-range systems received %a lot of
considerable attention in the last years, for reviews see for example \cite{RuffoRev} and \cite{defenu2021} for classical and quantum systems respectively. For what we are going to be concerned with in this paper, we remind that, at equilibrium, the universality class of a one-dimensional long-range system depends on the value of $\sigma$. Indeed, for $-1<\sigma<0$, the critical exponents are the mean field ones, that is the ones that we obtain by putting $\sigma=-1$. Then, there exists a non universal value $\sigma^{*}$ such that
%, for $0<\sigma<\sigma^{*}$ the critical exponents are either mean field or functions of $\sigma$, while
for $\sigma>\sigma^{*}$ we recover the critical exponents of the short-range case $\sigma=\infty$. Typically $\sigma^{*}>0$.
Furthermore, excitations in %these
long-range systems can propagate at diverging velocity 
\cite{Torcini1997,Metivier2014} and therefore we can expect some form of superdiffusive transport. There are already several, mainly numerical, studies of heat transport in long-range interacting systems that confirm these expectations. On the classical side, the heat transport 
was analyzed for the long-range XY model \cite{Olivares2016,iubini2018} the FPUT chain in \cite{Bagchi2017, iubini2018,di_cintio2019,wang2020thermal,bagchi2021heat}
and the lattice $\varphi^4$ theory \cite{iubini2022hydrodynamics}. In all cases Fourier's law is violated in different ways according to the value of $\sigma$. 
Scaling analysis of equilibrium correlations also suggests that hydrodynamics
is non-standard \cite{di_cintio2019,iubini2022hydrodynamics}. 
Thus, one may interpret transport as a fractional diffusion process 
with energy carriers performing L\'evy flights, with jump statistics
controlled by the exponent $\sigma$.

A classical harmonic long-range model with a stochastic dynamics was studied analytically in \cite{tamaki2020,suda2022superballistic} and the heat flux and temperature profile for a mean-field chain were computed in \cite{defaveri2021, ucci2022}. The same system was studied in the quantum regime in \cite{ucci2022} and a hydrodynamic approach to study transport in quantum magnets was proposed in \cite{schuckert2020}. %On the quantum side,
We refer again to \cite{defenu2021} for more references on the study of
dynamics and transport in quantum long-range systems.
However, in the literature there is not yet a detailed study of the plain
harmonic
chain with power-law interaction, and this contribution aims at filling this
gap.  We will show that 
the results are far from trivial in the strong long-range case and deserve 
careful analysis.

More precisely, in this paper we study numerically heat transport in a quadratic chain with a power-law interaction by coupling the  first and last site of the system to two heat baths at different temperature. We focus on computing the heat flux in the stationary state
with different approaches. 
In section \ref{sec1} we introduce the model and the main methods that
we will use to compute the heat flux. In section \ref{sec2}-\ref{sec4} we
report 
an analysis based on the spectral properties 
of the nonequilibrium Green's function and the transmission spectra and we discuss them. Finally, we draw our conclusions in section \ref{sec6}.

\section{Model and methods} \label{sec1}

\subsection{The long-range coupled harmonic chain}

We consider a one-dimensional chain of particles with a power-law interaction:
    \begin{equation} \label{ham}
        H=\frac{1}{2}\sum_{i}p_{i}^{2}+\frac{1}{2}\sum_{ij}x_{i}\Phi_{ij}x_{j}, 
    \end{equation}
where the interaction matrix $\Phi$ is given by:
\begin{equation} \label{phi}
    \Phi_{ij} = \left( 2\delta_{ij}-\frac{1}{N_{\sigma}} \frac{1}{|i-j|^{1+\sigma}}\right), \quad N_{\sigma}=\sum_{l=1}^{N} l^{-\sigma},
\end{equation}
where $ N_{\sigma}$ is the usual Kac factor introduced to guarantee
extensivity of the energy, chosen as site-independent. The matrix
correctly reduces to the discrete Laplacian for large $\sigma$. Note that definition \eqref{phi} corresponds to open boundary conditions, which are the ones appropriate for our problem due to the presence of the baths. For long-ranged systems we expect that the role of boundary conditions %will
can have very important consequences, even more than for short-ranged systems, 
%be relevant, but
and we focus on this natural choice for simplicity.

In the case of open boundary conditions the spectrum of matrix $\Phi$ is, to the best of our knowledge, 
not known analytically. The usual standing waves 
are not eigenvectors and the matrix cannot be diagonalized exactly.
Even in the continuum limit,  this would correspond to solving the 
spectral problem for the fractional Laplacian in a finite domain, 
which is notoriously not straightforward \cite{zoia2007fractional}.

For comparison, it is useful to recall the solvable 
case for periodic boundary condition where  the proper definition of $\Phi$ is:
 \begin{equation} \label{phi_pbc}
    \Phi_{ij} = \left( 2\delta_{ij}-\frac{1}{N_{\sigma}} \frac{1}{min(N-|i-j|^{1+\sigma},|i-j|^{1+\sigma})}\right).
\end{equation}
Here the spectrum  
is known, see for example \cite{defenuspectrum}. 
Due to translational invariance, the eigenvectors are plane waves 
of wavenumber $k$. The nature of the eigenfrequency spectra 
strongly depends on whether $\sigma$ is positive or negative. In the first case, the system has a proper continuum limit and for low momenta $k$ the squared frequencies $\omega^{2}$ of the plane waves behave as:
\begin{equation}
	\omega_{k}^{2} \approx
	\begin{cases}
	|k|^{\sigma}, \quad &0<\sigma<2, \\
	k^{2}, \quad &\sigma >2.
	\end{cases}
\end{equation}
Thus, for $\sigma>0$ one has the standard acoustic dispersion and
a finite group velocities while the group velocity diverges as $|k|^{\frac{\sigma-2}{2}}$
in the first case. This result can 
also be 
derived from the continuum limit,  corresponding to a fractional wave equation in the infinite domain \cite{Tarasov2006}.
On the other hand, if $\sigma<0$ the spectrum remains discrete even in the thermodynamic limit and contains a countable infinite number of frequencies that accumulate at the band edge \cite{defenuspectrum}.

To simulate the non-equilibrium steady state, we follow the usual procedure and
%attach
connect the first and last sites of the system to two Langevin 
heat baths at temperatures $T_{L}$ and $T_{R}$, respectively. The coupling with the baths introduces both noise and dissipation in the dynamics of the system. The resulting equations of motion are:
\begin{equation} \label{eom}
    \ddot{x}_{i} = -\sum_{j} \Phi_{ij} x_{j} + \delta_{i1}\left(\xi_{L}-\lambda \dot{x}_{i}\right) + \delta_{iN}\left(\xi_{R}-\lambda \dot{x}_{i}\right),
\end{equation}
where the $\xi$'s are Gaussian noises that satisfy the fluctuation-dissipation relation:
\begin{equation}
    \Braket{\xi_{a}(t)\xi_{a}(t')} = 2 T_{a} \lambda \delta(t-t'), \quad a=L,R.
\end{equation}
After a transient, the system reaches a stationary state: we are interested in the heat flux of the chain in this state. To compute this quantity, we will employ three different methods. 

\subsection{RLL approach}

The first method was introduced long time ago 
in this context in %by Rieder, Lebowitz and Lieb
\cite{RLL}. It consists in solving the many-body Fokker-Planck equation related to \eqref{eom} (in the following we will refer to this method as the RLL method). In particular, 
defining the vector $y=(x_{1},...x_{N},p_{1},...p_{N})$, and denoting by $P(y,t)$ its 
probability at time $t$, 
the aforementioned equation reads as:
\begin{equation}
\frac{\partial P(y,t)}{\partial t} = A_{ij}\frac{\partial}{\partial y_{i}}(y_{j}P)+\frac{1}{2} D_{ij}\frac{\partial^{2}P}{\partial y_{i}\partial y_{j}} , \quad \label{fp} 
\end{equation}
where the drift and diffusion matrices are
\begin{equation}
A= \begin{pmatrix} \mathbb{O} & -\mathbb{I} \\ -\Phi & \lambda \mathcal{R} \end{pmatrix} , \quad D=
\begin{pmatrix} \mathbb{O} & \mathbb{O} \\ \mathbb{O} & 2k_{B}\lambda T(\mathcal{R}+\eta \mathcal{S}) \end{pmatrix},
\label{mats}
\end{equation}
where
\begin{align}
&\begin{cases}
T=\dfrac{T_{L}+T_{R}}{2}, \\ \eta=\dfrac{T_{L}-T_{R}}{T}
\end{cases}, \quad 
\mathcal{R}_{ij} = \delta_{ij}(\delta_{i1}+\delta_{iN}), \quad\\ &\mathcal{S}_{ij}=\delta_{ij}(\delta_{i1}-\delta_{iN}). \label{mats1}
\end{align}
The solution of equation \eqref{fp} is a multi-variate Gaussian whose covariance matrix is given by the matrix of correlations among the canonical coordinates:
\begin{align} 
                                &P(y,t) \propto \exp{\left\{-\frac{1}{2}C^{\,-1}_{ij}y_{i}y_{j}\right\}}, \quad C=\begin{pmatrix} \Braket{x_{i}x_{j}} & \Braket{x_{i}p_{j}} \\ \Braket{p_{i}x_{j}} & \Braket{p_{i}p_{j}} \end{pmatrix}. \label{gauss}
\end{align}
By plugging \eqref{gauss} in the Fokker-Planck equation \eqref{fp} we get:
\begin{equation} 
\partial_{t}C = D-AC-CA^{T}.
\end{equation}
Furthermore,  in the stationary state $\partial_{t}C=0$, so we get the so-called 
(continuous) Lyapunov equation:
\begin{equation}
AC+CA^{T}=D,
\end{equation}
which %can
has %then
to be solved numerically. Knowing the various correlators, we can then express the heat flux in the stationary state as the difference between the temperature of the left bath and the temperature of the first site:
\begin{equation}
    \mathcal{J}=\lambda \left(T_{L}-T_{1}\right), \quad T_{i} = \frac{1}{2}\Braket{p_{i}^{2}}. \label{hr_rll}
\end{equation}

\subsection{Nonequilibrium Green's function}

The second method consists in writing the exact solution to \eqref{eom} in terms of the Green's function $G(\omega)$, which is possible due to the linearity of the equations. The details of this method are explained in refs. \cite{dhar2006, dhar2016heat, DHARREV}. Since we are interested in the stationary state, we work directly in frequency space:
\begin{eqnarray}
&& \tilde{x}_{l}(\omega) = \sum_{ln} G_{ln}(\omega)(\tilde{\xi}_{L,n}(\omega)+ \tilde{\xi}_{R,n}(\omega)),\\ 
&& G(\omega)=\left(-\omega^{2}\mathbb{I}+\Phi+i\lambda\omega \mathcal{R}\right)^{-1},
    \label{green}
\end{eqnarray}
where the tilde indicates the Fourier transform and $\mathcal{R}$ is the matrix defined in Eqs.(\ref{mats}, \ref{mats1}).
%is a matrix that has nonzero entries equal to $1$ only on the sites coupled to the baths: $R_{ij}=\delta_{ij}(\delta_{i1}+\delta_{iN})$.
%\textcolor{red}{e' la stessa gia' definita sopra?!}
%Then,
As explained in \cite{DHARREV}, we can express the heat flux in the stationary state as:
    \begin{equation}
        \mathcal{J}=\frac{2\Delta T \lambda^{2}}{\pi}\int_{0}^{\infty} d\omega \,  \omega^{2} |G_{1N}(\omega)|^{2}. \label{hf_green}
    \end{equation}
    
\subsection{Generalized eigenvalue method}
    
There is in the literature another approach to the Green's function method, called \textit{generalized eigenvalues method}, which we briefly outline below (for a more detailed explanation see \cite{freitas2014, freitas2016, tisseur2001}). Let $G^{L}(s)$ be the Green's function defined in Laplace's space:
\begin{equation} \label{green_lap}
G^{L}(s) = \left(-s^{2}\mathbb{I}+\Phi+\lambda s R\right)^{-1},
\end{equation}
and introduce the $2N$ complex numbers $\{s_{a}\}_{a=1}^{2N}$ and the $2N$ vectors $\{\bm{r}_{a}\}_{a=1}^{2N}$ as defined by the following linear problem:
\begin{equation}
	G^{L}(s_{a})\bm{r}_{a} = 0.
\end{equation}
Then, the Green's function \eqref{green_lap} can be written as \cite{tisseur2001}:
\begin{equation} \label{green_poles}
	G^{L}(s) = \sum_{a=1}^{2N} \dfrac{s_{a}}{s-s_{a}} \bm{r}_{a} \bm{r}_{a}^{\dagger}.
\end{equation}
Note that the $s_{a}$ come in complex conjugate pairs.
We now recall that we can obtain the Green's function in frequency space $G(\omega)$ via a Wick rotation $G(\omega) = G^{L}(-is)$. Then we can compute the integral in \eqref{hf_green} with a contour integration one finding \cite{freitas2014}:
\begin{equation} \label{hf_poles}
	\mathcal J = 2 \Delta T \lambda^{2} \sum_{a,b=1}^{2N} \dfrac{s_{a}^{3} s_{b}}{s_{a}+s_{b}} r_{a,1} r_{a,N}^{\*}r_{b,N} r_{b,1}^{\*}.
\end{equation} 
Formula \eqref{hf_poles} gives yet another way of computing the heat flux and extract the scaling exponents.

\subsection{%Green's function method
Comments}

Before proceeeding, let us comment on the numerical issues connected with the
above approaches.
%In particular,
The numerical implementation of the RLL method is rather straightforward resorting to the numerical routines available to solve the Lyapunov equation based on the Bartels-Stewart algorithm, as implemented 
for instance in the SciPy library \citep{2020SciPy-NMeth}. 
Indeed, %we
one can easily reach sizes of $N\sim 10^{3}$.
Some convergence issues may arise in the case of strong degeneracies \cite{ucci2022}. 
The numerical implementation of the Green's function method can be more involved than the one of the RLL method. Indeed, we need to numerically invert the matrix in the definition of the Green's function \eqref{green} in the range of $\omega$ where the 
transmission is non-vanishing in order to be able to compute the integral in \eqref{hf_green}. Furthermore, the sampling over $\omega$ has to be fine enough
to ensure accuracy, especially if the transmission coefficient oscillates 
rapidly. This difficulty does occur in our model, as it will   
be clear in what follows. In practice, it is difficult to study 
lattices larger than  $N\sim 10^{2}$ using this method. 
The generalized eigenvalues method has the advantage of reducing the 
problem to the calculation of the eigenvalues and eigenvectors of a
$2N\times 2N$ matrix \cite{freitas2016}, which can be done by standard linear algebra 
routines, the main limitation being memory storage and accuracy 
of very small eigenvalues and avoiding the sampling problem.

\section{Heat Flux} \label{sec2}

In the short-range case, $\sigma=\infty$, two of the methods outlined above have been used to obtain exact analytical results for the heat flux in the thermodynamic limit \cite{RLL, DHARREV}. This is possible because the matrix of the interactions $\Phi$ reduces to the discrete Laplacian, which is a tridiagonal matrix. In our case the matrix $\Phi$ is dense, and we are unable to either solve analytically the Lyapunov equation or to exactly compute the Green function. Nonetheless, it is possible to obtain a certain amount of informations about the heat flux numerically.

\subsection{Small coupling}

If the coupling with baths $\lambda$ is small, a perturbative calculation 
of the steady-state current is possible in terms of the eigenvalues and 
eigenvectors of the isolated harmonic chain. 
This approach yields the so-called Matsuda-Ishii's formula,  whereby $\mathcal J\approx\mathcal{J}_{MI}$ to the leading 
order in the coupling constant \cite{Matsuda1970, Lepri_review}, with
$\mathcal{J}_{MI}$  given by 
\begin{equation}
\mathcal{J}_{MI}=\lambda \Delta T\sum_k
\frac{\psi^2_{k,1}\psi^2_{k,N}}{\psi^2_{k,1}+\psi^2_{k,N}}
\end{equation}
where $\Delta T=T_L-T_R$ and 
$\psi_{k,,n}$ denotes the $n$ component of the $k$th eigenvector
of the matrix $\Phi$ defined in (\ref{phi}).
For the model we consider 
here (which is homogeneous and mirror-symmetric,  i.e. the first and last component of each eigenvector are equal $\psi_{k,1}=\psi_{k,N}$ for $k=1...N$) the above 
expression simplifies to
\begin{equation} \label{mi}
    \mathcal{J}_{MI}=\dfrac{\lambda \Delta T}{2} \sum_{k} \psi_{k,1}^{2} =\dfrac{\lambda \Delta T}{2},
\end{equation}
where in the last step we used the property of completeness of the set of eigenvectors.
Note that eq. \eqref{mi} expresses the fact that the chain is a ballistic conductor. 

\begin{figure}[th]
    \includegraphics[scale=0.23, trim =1.5cm 0 0 0]{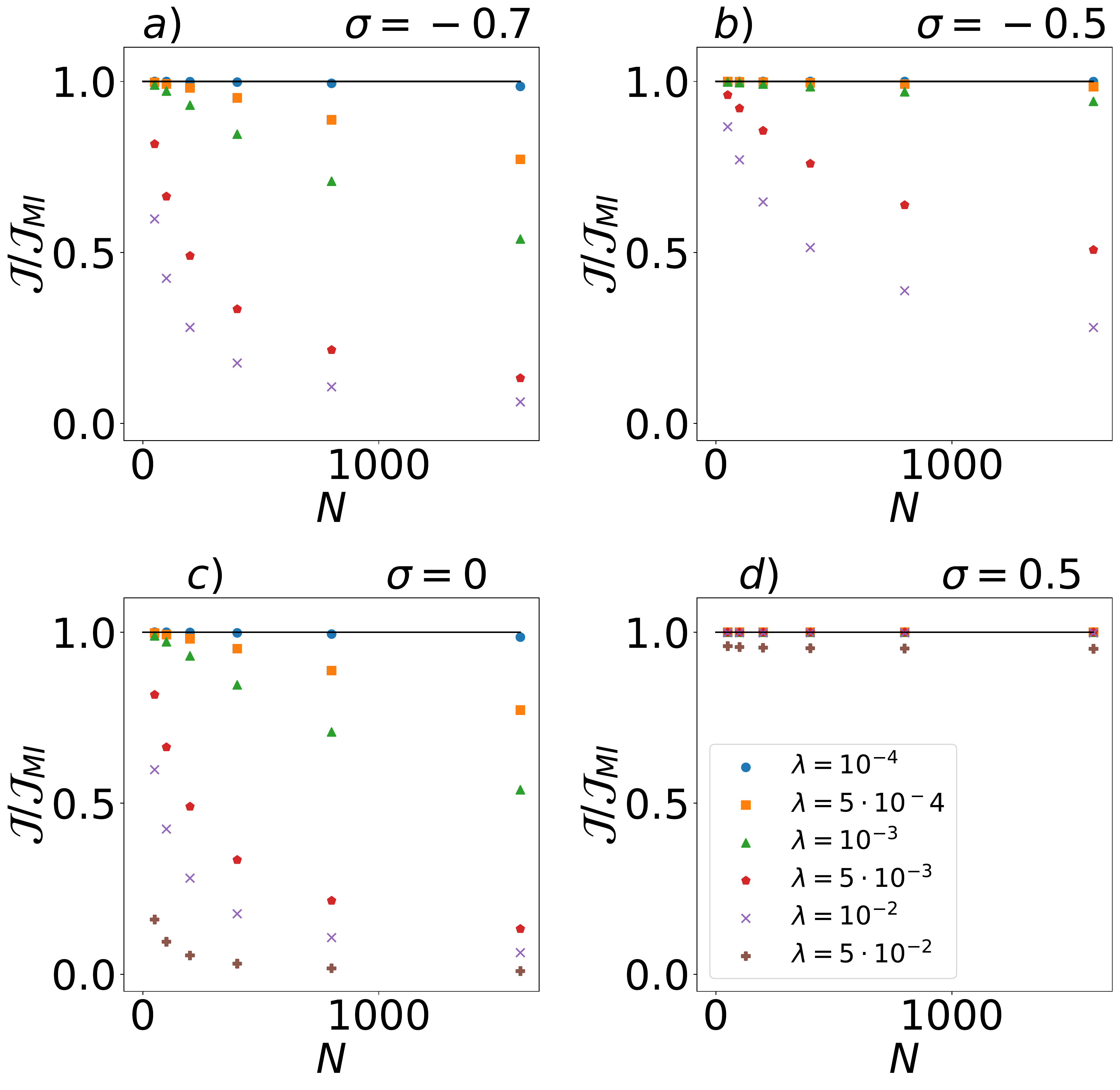}
\caption{Plots of the ratio between the heat flux $\mathcal{J}$, computed numerically with the RLL method, and the Matsuuda-Ishii heat flux \eqref{mi} versus the system size $N$ for several values of $\sigma$ and $\lambda$ in the weak coupling regime.} \label{fig:fig1}
\end{figure}

Typically, in the short-range case $\sigma \rightarrow \infty$, this result
applies for $\lambda \ll \lambda_{0} \approx \mathcal{O}(1)$. In the our long-range case, however, the situation is more complicated. In Fig. \ref{fig:fig1}, we compare formula \eqref{mi} and the numerical solution of the Lyapunov equation. As we can see, \eqref{mi} holds for $\lambda$ smaller than a certain threshold $\lambda_{0}(\sigma,N)$, that depends both on $N$ and on $\sigma$. More specifically, $\lambda_{0}$ decreases with $\sigma$ and with $N$. On the other hand, for $\sigma>0$ the perturbative approximation
holds well in the considered range.

\begin{figure}[th]
    \includegraphics[scale=0.16,clip]{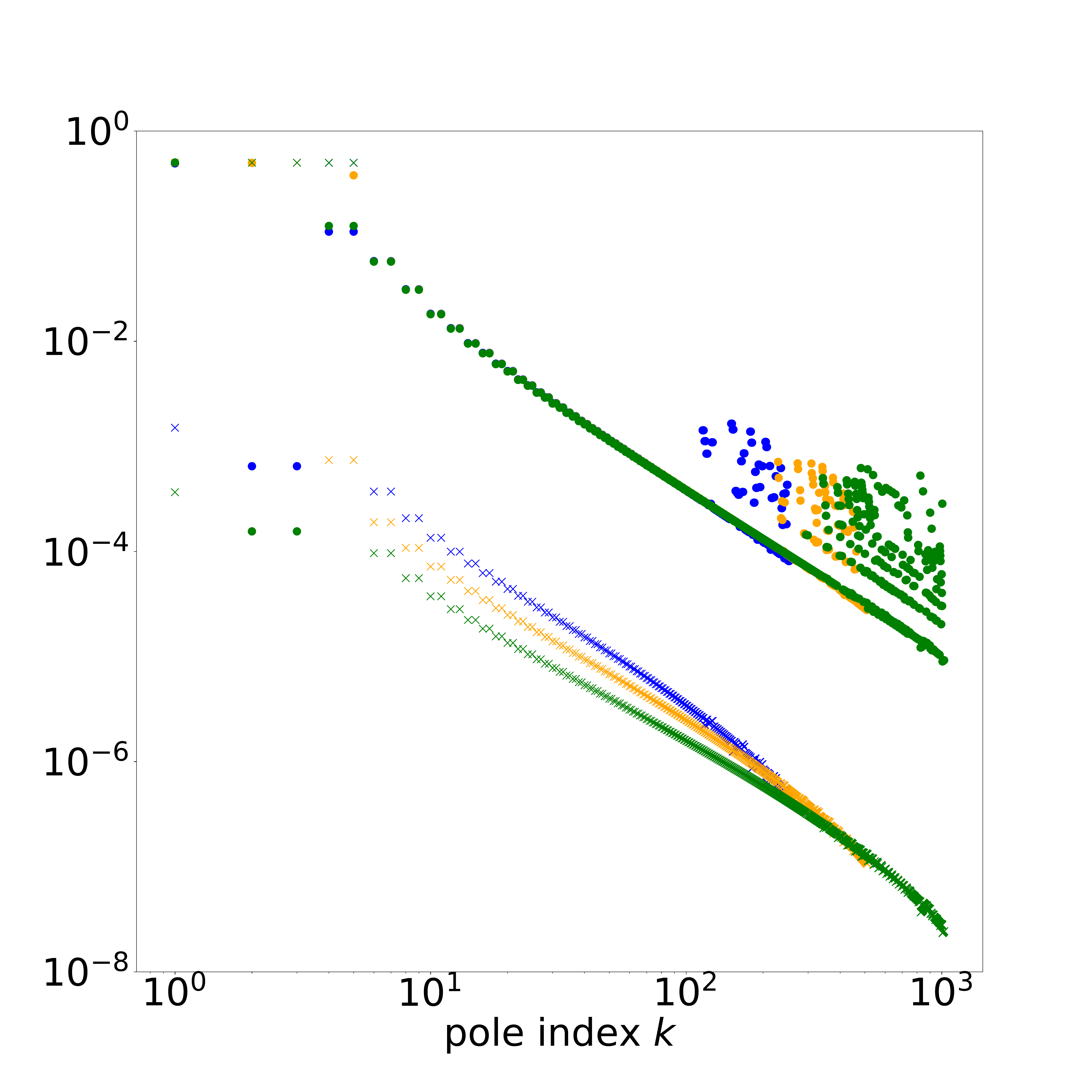}
\caption{Plots of the spacing between the imaginary parts of the poles of the Green's function $Im(s_{k+1})-Im(s_{k})$ (circles) and the real parts of the poles $Re(s_{k})$ (crosses) for $\sigma=-0.5$. Different colors correspond to different system's size: $N=256,512,1024$ in blue, orange, green, respectively.}
\label{deltaw}
\end{figure}

To have some insight into these deviations we may perform some further checks.
Usually the perturbative approach is justified assuming that the separation 
of the unperturbed normal mode frequencies is smaller than the 
typical dissipation caused by the coupling with the baths \citep{freitas2014}.
This assumption can actually be checked by examining the poles $s_a$. 
In particular, we compare the 
spacings between the imaginary parts of consecutive poles $Im(s_{a+1}-s_{a})$ and the 
real parts  $Re(s_{a})$ 
. As we can see from Fig. \ref{deltaw}, the former
is always much larger than the latter, therefore  
this assumption is justified.
This suggests that the observed deviations from the Matsuda-Ishii formula may have 
a different origin.

\subsection{Strong coupling}

\begin{figure}
    \includegraphics[scale=0.19, trim = 4cm 0 4cm 2cm]{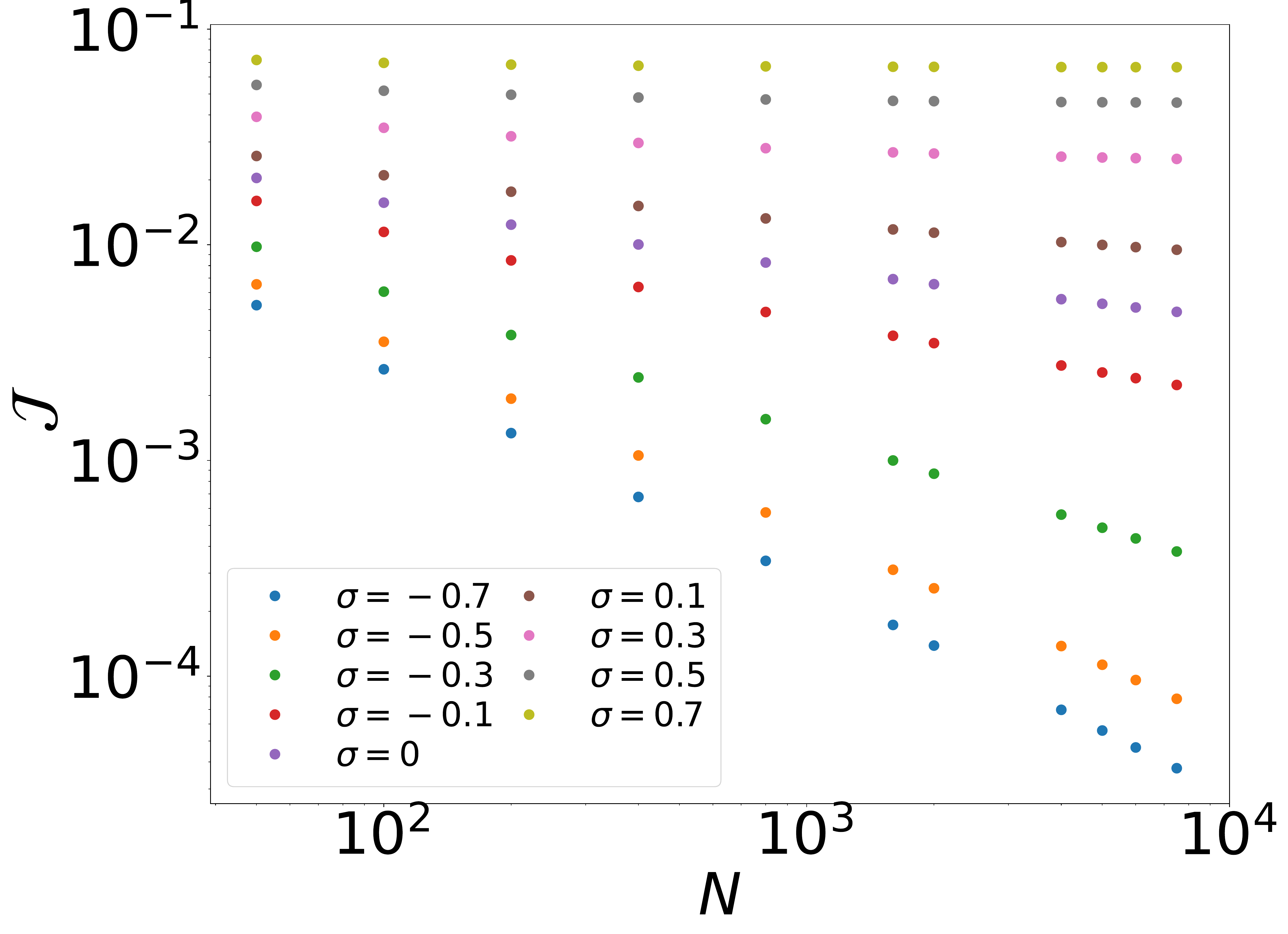}
    \caption{Log-log plot of the heat flux $\mathcal{J}$ versus the system's size $N$
    for $\lambda=1$ and different values of the long-range exponent $\sigma$. 
    The  flux is computed using the RLL method as described in the text.}
	\label{plot_flux}
\end{figure}

We now want to understand how the flux scales with the system size $N$ for not
too weak coupling 
$\lambda$. In order to so, we computed the heat flux using the RLL method for several values of $N$ and $\sigma$ for $\lambda=1$ (and we will set $\lambda=1$ for the rest of the paper)
As shown in Fig.\ref{plot_flux} the data can be fitted with a power law $\mathcal{J} \propto N^{-\gamma}$.

\begin{figure}[th]
    \includegraphics[scale=0.21,trim = 8cm 0 0 0  ]{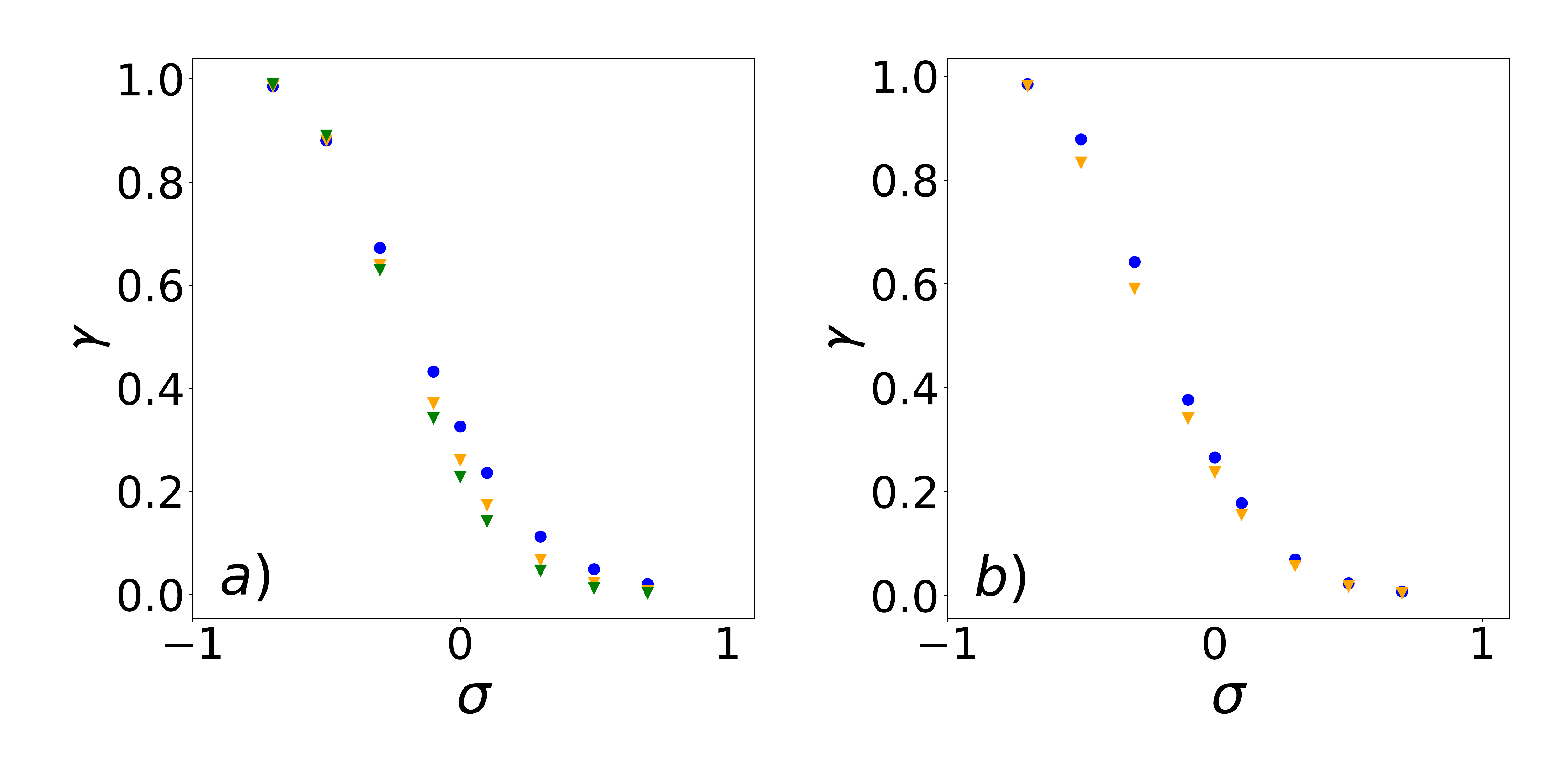}
    \caption{Plot of the scaling exponent of the flux $\gamma$, defined as $\mathcal{J} \propto N^{-\gamma}$, $(a)$, we report the exponents obtained by fitting a power law 
    on the  heat flux obtained with the RLL method. To check the finite-size 
    effects, each data set corresponds to a fit over different length ranges,  
    $50\le N \le 1600$ (circles), $500 \le N \le 2000$ (squares), $1500 \le N \le 7500$ (triangles). Panel $(b)$, comparison between the exponents obtained by the RLL method (circles) and the generalized eigenvalue method (triangles). 
    }
    \label{exponents}
\end{figure}

Although the direct computation of the Green's function is numerically cumbersome, we can easily compute its poles, compute the heat flux according to \eqref{hf_poles} and fit a power law as we did before. In panel $b)$ of Fig. \ref{exponents} we report both the exponents fitted with the generalized eigenvalues method and with the RLL method. As we can see, they are qualitatively in agreement. 

The results of fits using the two methods are reported in Fig. \ref{exponents}. 
We can identify three regions. The region close to the mean-field case $\sigma=-1$ and the one close to the short-range case $\sigma>1$, where finite-size effects are almost absent, and an intermediate region in which finite-size effects are quite strong. We also note that $\gamma$ seems to be %correctly
converging to the short-range value $\gamma=0$ while $\sigma$ goes to $1$. Summarizing, even if we are not able to extract the exact values of the exponents, it is clear that the flux scales with some nontrivial power of the system's size $N$. \\

\section{Transmission spectra}
\label{sec3}

\begin{figure*}[ht]
	\includegraphics[scale=0.155,clip]{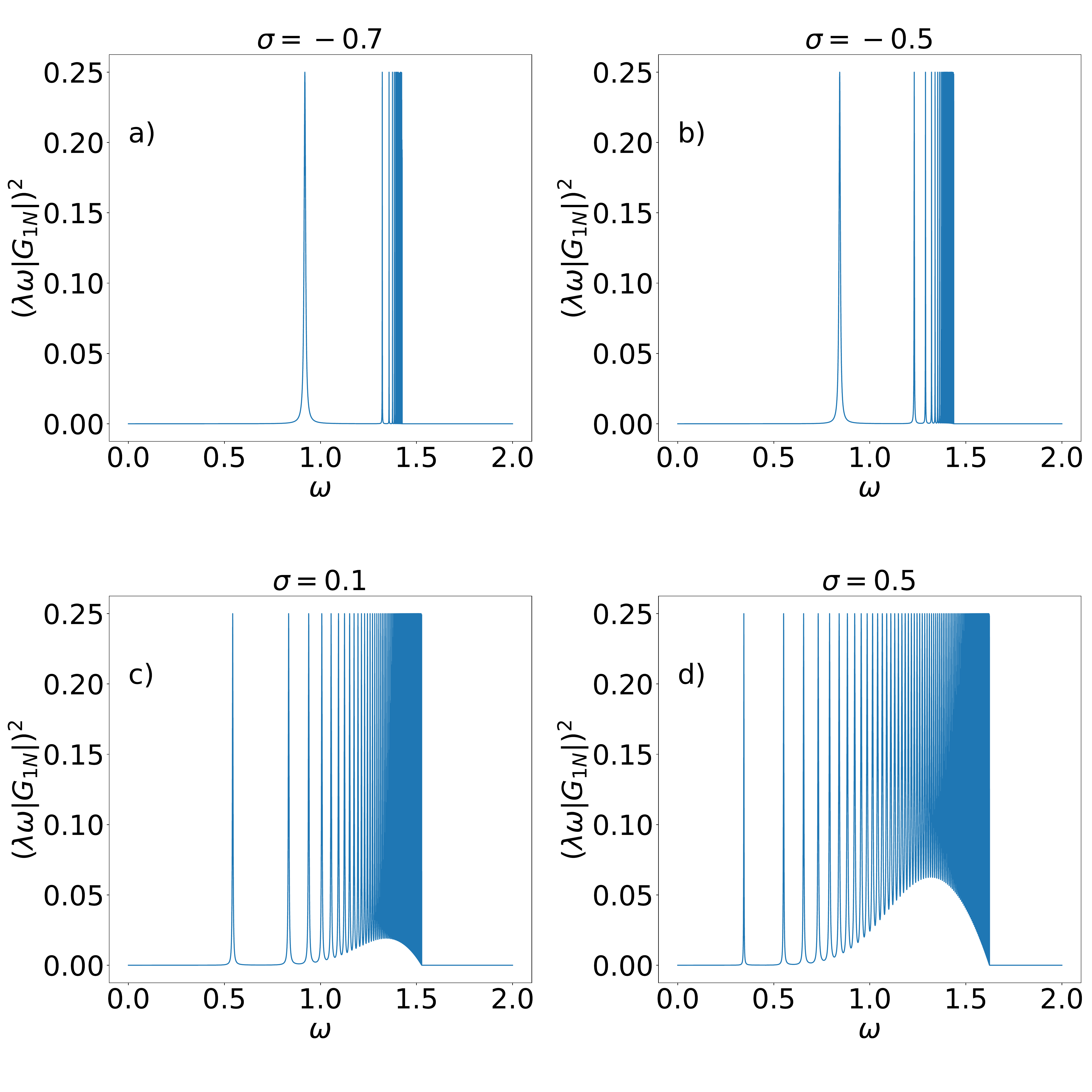}
	\includegraphics[scale=0.17, clip]{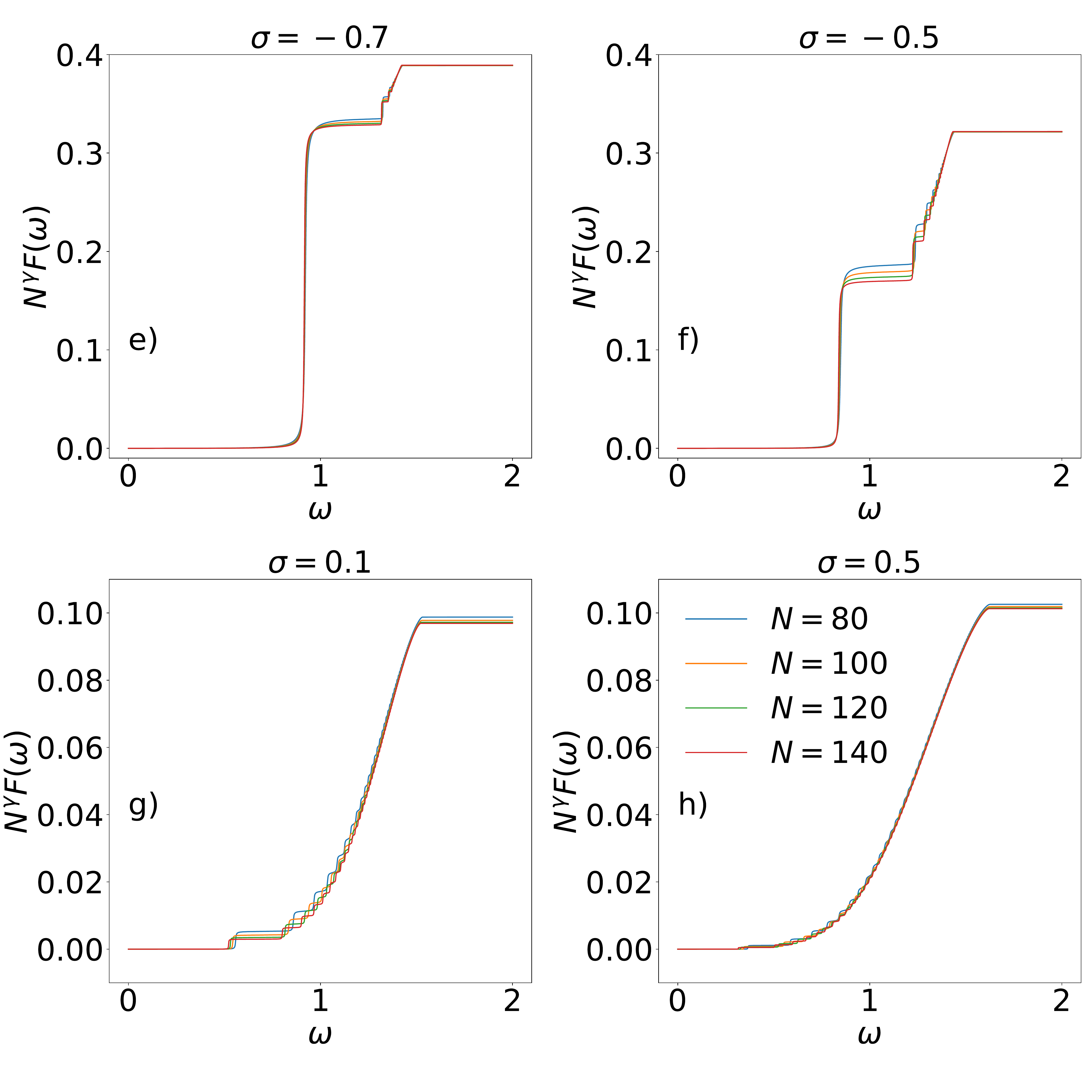}
	\caption{Panels $a)$, $b)$, $c)$, $d)$ : transmission spectra (the integrand of the heat flux 
	expression \eqref{hf_green}) for different values of the range exponent 
	$\sigma=-0.7,0.5,0.1,0.5$ and for a chain with $N=100$.
	Only positive frequencies are reported.
 Panels $e)$, $f)$, $g)$, $h)$: rescaled cumulative function $N^{\gamma}F(\omega)$, for $N=80,100,120,140$ and $\sigma=-0.7, -0.5, -0.1, 0.5$ in panels $a)$, $b)$, $c)$, $d)$ respectively. The values of $\gamma$ are taken from the blue points in Figure \ref{exponents}. The abrubt increase of the cumulative function in panels $e)$ and $f)$ at $\omega\approx 1.3$ is due to the dominant contribution of the first peak in panels $a)$ and $b$. The subsequent, smaller, jumps are due to the contributions of the other peaks.	
}
	\label{peaks}
\end{figure*}

To understand the origin of the nontrivial dependence of the flux on the 
size, let us investigate the transmission spectrum of the chain. 
We begin by plotting the transmission coefficient, namely the integrand in \eqref{hf_green} as a function of the frequency $\omega$. In Fig. \ref{peaks} we report its plot for several values of $\sigma$. We can see that it is characterized by a rather complicated peak structure which consists of $N-2$ peaks (as can be checked numerically).

A manin point we want to make and explore is that the structure of such resonances determines the scaling of the current. Notice that a change of sign in $\omega$ in \eqref{green} is equivalent to the complex conjugation of $G(\omega)$. Since the transmission coefficent depends on the square modulus of $G(\omega)$ it is an even function of $\omega$ and we can therefore restrict ourselves to study positive frequencies.
Let us denote by $\omega_k$, $k=1,2\ldots$ the location of the
peak frequencies for positive $\omega$. 
The peaks accumulate at a band-edge frequency ${\omega_B}<2$, i.e 
$\omega_k\to {\omega_B}$ for $k$ large. Furthermore, upon approaching ${\omega_B}$, the width of the peaks decreases. Notice that this is the reason why it is important to finely sample the Green's function in $\omega$, especially in the proximity of the band edge. Indeed, we used a logarithmic sampling in order to increase the sampling points near 
${\omega_B}$. The integrand is thus a much more complicated function of $\omega$ with respect to the mean-field case $\sigma=-1$ \cite{defaveri2021, ucci2022}, where only the first peak is present. It can be checked numerically that the first few peaks are Lorentzian with amplitude $\Delta_k \approx N^{-1}$, exactly like the peak in mean-field case. The subsequent peaks are too narrow to be resolved. For positive values of $\sigma$ the situation becomes even more complicated, as a curve emerges below the peaks, as we can see in Fig. \ref{peaks} for $\sigma=0.5$. 

For the reasons outlined above, it seems more convenient to consider the cumulative function $F(\omega)$, that is, the integral \eqref{hf_green} performed up to frequency $\omega$. In the rightmost panels of Fig. \ref{peaks} we report the function $F(\omega)$ for several values of $N$ of order $10^{2}$ and $\sigma$, rescaled by $N^{\gamma}$, where $\gamma$ is the exponent obtained with the RLL method for values of $N$ of order $10^{2}:10^{3}$. As we can see, the curves nicely collapse for $\sigma=-0.7,-0.5$, but for higher values of $\sigma$, such as $\sigma=-0.3$, the collapse is not as good due to the finite-size effects, as expected. Regardless of the lack of further quantitative progress in the computation of the exponents, the qualitative information about the peak structure will be crucial in our understanding of the model, as we will see later.

\section{Poles of the Green's function}
\label{sec4}

In view of the numerical difficulties encountered above and for comparison, we 
also performed a study of poles of the Green's function. These are computed through the 
generalized eigenvalue method described above.

The main advantage of the analysis is that we gain a new perspective on the peak structure discussed before. Indeed, the positions $\omega_k$ of the peaks in Fig. \ref{peaks} are given by the absolute value of the imaginary part of $s_{a}$, while the absolute value of the imaginary part should be proportional to their widths $\Delta_k$. 

\begin{figure*}[ht]
	\includegraphics[scale=0.18, trim = 5.5cm 0 0 0]{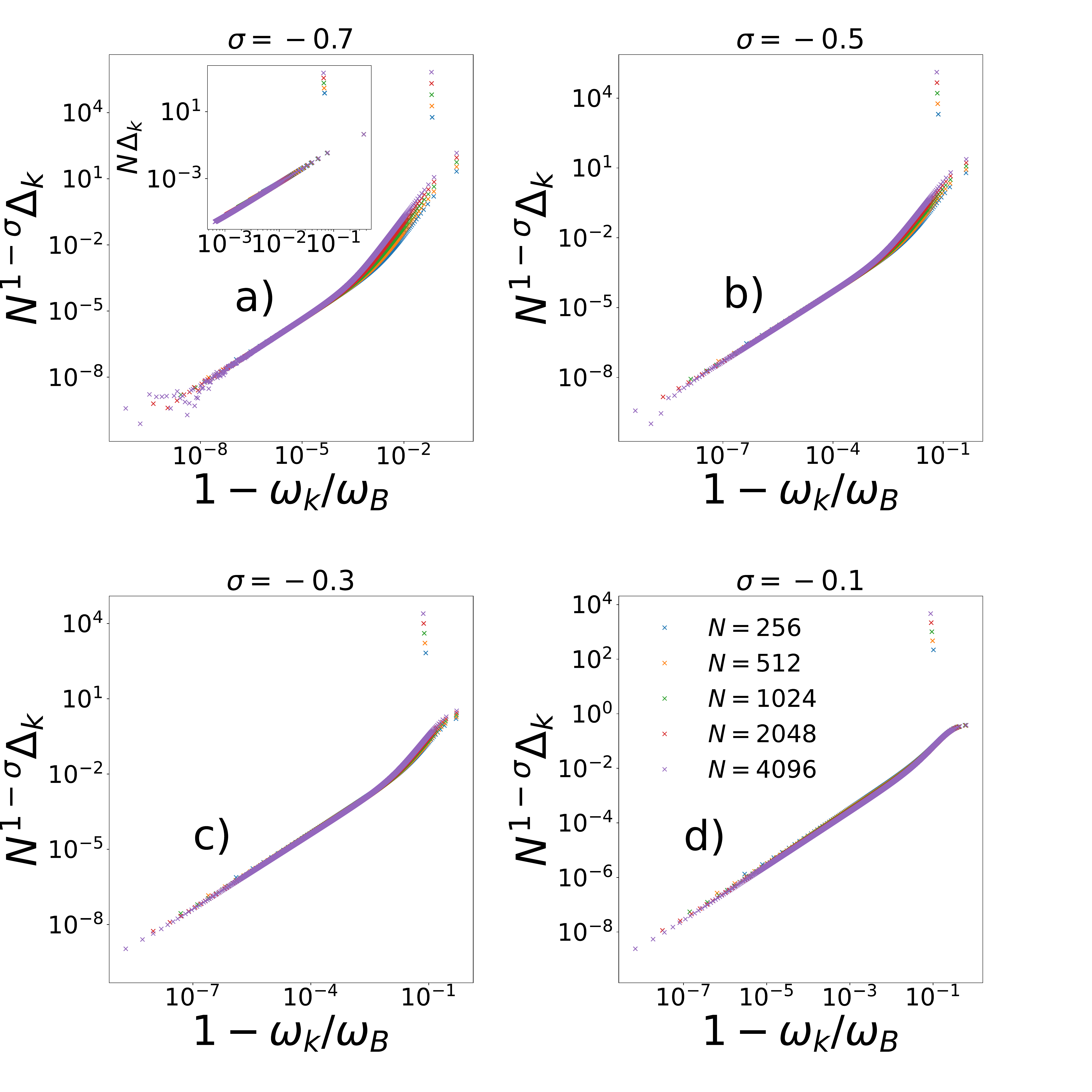}
	\includegraphics[scale=0.19, trim = 0 0 10cm 0]{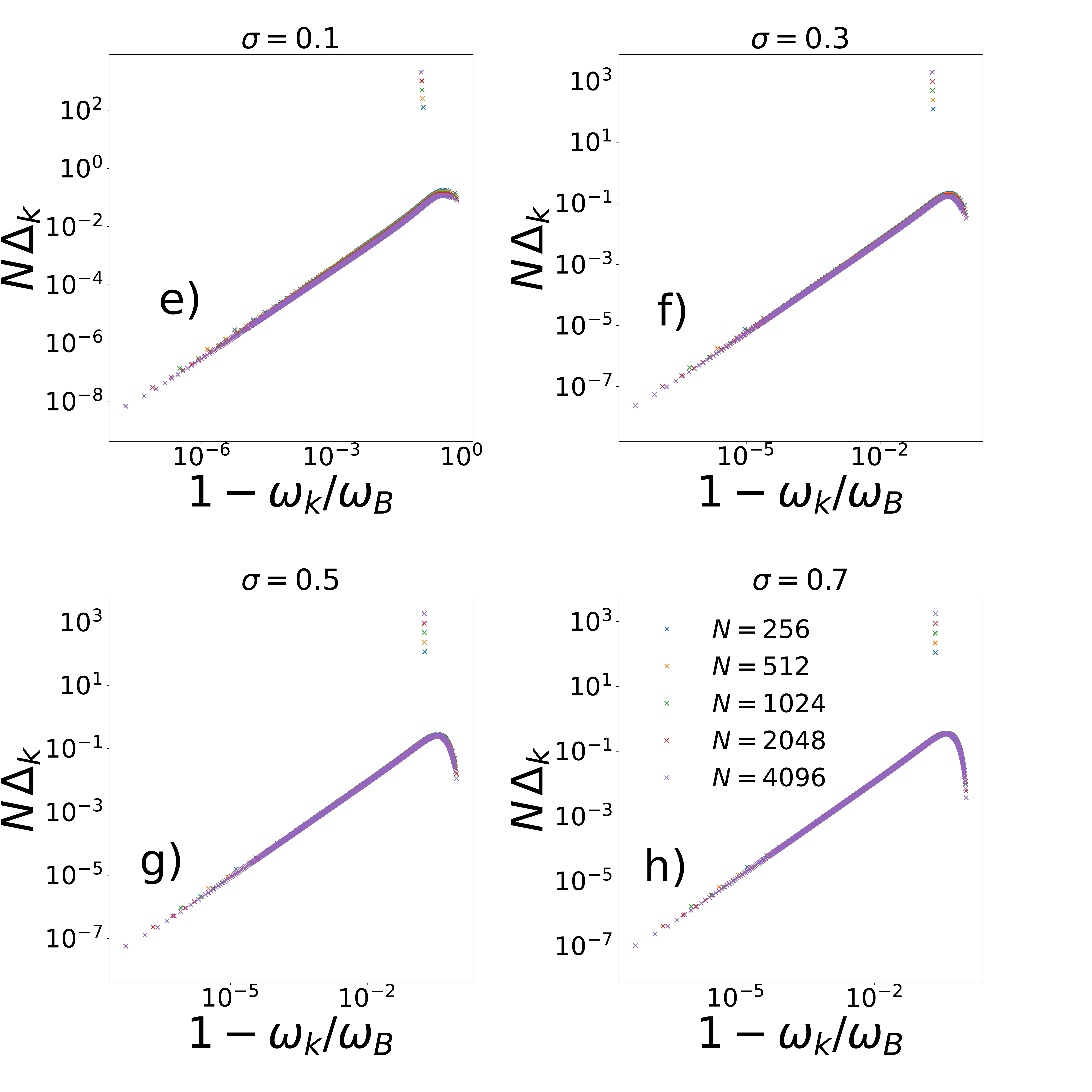}
	\caption{Real parts of the poles of the Green's functions $s_{a}$ 
	versus the distance of their imaginary parts from the band-edge. 
	Leftmost panel: $\sigma <0$, vertical axis rescaled by $N^{\delta}$ 
	with $\delta \approx 1-\sigma$.
	The inset in panel (a) demonstrates the different scaling the collapse for the widths of the first peaks (the first $80,160,320,640,1280$ for $N=256,512,1024,2048,4096$, respectively) rescaled by $N$. For the other values of $\sigma$, we get the same scaling for the first peaks. Rightmost panel: same for $\sigma > 0$, with vertical axis rescaled by $N$.
	Note that such scaling works for the whole spectrum in this case.} \label{widths_neg}
\end{figure*}

In particular, we consider all the peaks as Lorentzian --
for simplicity, but also because all the peaks that we were able to resolve
are actually very well approximated by a  Lorentzian -- with width given by $\Delta_{k}(N)= Re(s_{k})$. In this approximation, as far as scaling with the size is 
concerned, the heat flux can be estimated as the sum of the widths of the 
peaks $\Delta_{k}(N)$. Furthermore, the height of each peak can be shown to be equal to $\lambda^{2}/4$ (indeed, note that in Fig. \ref{peaks}, in which $\lambda=1$, the heights of the peaks are all the same and equal to $1/4$). Thus, we replace the integrand in eq. \eqref{hf_green} with a sum of normalized Lorentzians, and we get:
\begin{equation} 
\label{hf_sum}
	\dfrac{\mathcal{J}(N)}{\Delta T} \approx \int_{-\infty}^{\infty} \dfrac{d\omega}{\pi} \sum_{k=1}^{N-2} \dfrac{\lambda^{2} \Delta_{k}(N)^{2}/4  }{(\omega-\omega_{k})^{2} +\Delta_{k}(N)^{2}  } =   \frac{\lambda^{2}}{4} \sum_{k=1}^{N-2} \Delta_{k}(N).
\end{equation} 
The relevant information should thus be contained in the dependence
of the  $\Delta_{k}$ on $k$ and $N$. Physically, this is the 
effective damping of plane waves due to the coupling with the thermal
reservoirs. 

The dependence of $\Delta_{k}$ on $N$ is reported in Fig. \ref{widths_neg}, 
where we plot (parametrically) the real parts of the poles as a function
of the imaginary ones, 
for negative  and positive values of $\sigma$, respectively. 
Since the resonances accumulates at the band-edges, it is convenient to 
report the frequencies as a function of their relative distance from $\omega_B$.
Let us focus on the case of negative $\sigma$, to begin with. 
From the leftmost panels of Fig. \ref{widths_neg}, it is seen that the poles can be grouped in two 
sets, each having different dependencies on $\omega_k$ and $N$.
Empirically, this is accounted for by the following scaling:
\begin{equation} \label{widths_scalings}
	\Delta_{k}(N) \approx 
	\begin{cases}
		d_{k}/N, \quad k<k_{o} \\ 
		d_{k}/N^{\delta}, \quad k>k_{o},
	\end{cases}
\end{equation} 
where $k_{o}<<N$ and $d_{k}$ do not depend $N$. We do not 
have an a-priori theoretical estimate of $\delta$, but we find that there is a good collapse upon choosing $\delta\approx 1+|\sigma|$. 
It is interesting to point out that the exponent $\delta$ can be interpreted as the fractal dimension of area below the graphs in Fig. \ref{peaks}. Indeed, if we increase the system's size $N$ new peaks emerge with progressively shrinking area and, in a putative $N\rightarrow \infty$ limit we would have an infinite number of peaks with vanishing area.

In addition, there are a few poles  whose widths do not follow this scaling and 
fall consistently well outside the collapsed curve. It actually turns out that there are two degenerate eigenvalues between the $s_{a}$s that do not follow the scaling law. However this is inconsequential, as one can check that the contribution of the these eigenvalues to \eqref{hf_poles} vanishes. Heuristically, this is because, as one can check, the eigenvectors related to these eigenvalues are localized at the endpoints of the chain and therefore do not contribute to transport. This also explains why the peaks in Fig. \ref{peaks} are $N-2$ instead of $N$.
We can therefore infer the following scaling law for the heat flux \eqref{hf_poles} plugging \eqref{widths_scalings} into \eqref{hf_sum}:
\begin{equation}
  \mathcal{J} \approx \dfrac{\sum_{k=1}^{k_{o}} d_{k}}{N} + \dfrac{\sum_{k=k_{o}}^{N} d_{k}}{N^{\delta}} %\approx
  \propto N^{1-\delta}.
\end{equation}
The first term scales as $N^{-1}$, since $k_{o}$ does not scale with $N$ (as can be inferred from Fig.\ref{widths_neg}). On the other hand, the second term scale as $N^{1-\delta}$ since each $d_{k}$ is of order $1$ and thus their sum scales as $N$. Finally, since $\delta>0$, we get the reported scaling for the heat flux.
For positive $\sigma$, the scaling of $\Delta_{k}$ is reported in the right-most 
panels of Fig. \ref{widths_neg}: as we can see in this 
case $\Delta_{k} \approx N^{-1}$, over the entire spectrum. Therefore, 
the estimate  the heat flux yields
\begin{equation}
	\mathcal{J} \approx \dfrac{\sum_{k=1}^{N} d_{k}}{N} \approx \mathcal{O}(1).
\end{equation}
So the heat flux for positive $\sigma$ behaves as the heat flux for $\sigma=\infty$ (the nearest-neighboors case), that is, it does not scale with $N$.

To summarize, according to approximation (\ref{hf_sum}) 
and the numerical estimate of $\delta$ extracted from the 
data, we find that the heat flux %should
scale as:
	\begin{equation} \label{gamma}
	  \mathcal{J} \propto N^{-\tilde{\gamma}}, \quad \quad \tilde{\gamma}
          %=
          \approx
		\begin{cases}
			1-\delta, \quad 	\sigma<0, \\
			0, \quad	\sigma>0.
		\end{cases}
	\end{equation}

As we already mentioned, see Fig. \ref{widths_neg}, we found a good collapse of the imaginary part of the poles of the Green's functions for $\delta \approx 1-|\sigma|$.
So this %would
yields
\begin{equation}
  \tilde{\gamma}\approx -\sigma
  \end{equation}
for negative $\sigma$. Admittedly, this estimate accounts only qualitatively for the 
behavior of the exponents as given in Fig. \ref{exponents}. The deviations
are sizeable and, in addition the  dependence of $\gamma$ on $\sigma$ appears 
to be non-linear. 
While this could be due to the aforementioned finite-size effects, the discrepancy is present even for values of $\sigma$ for which the exponent $\gamma$ has basically converged (for example $\sigma=-0.7, -0.5$). Another possibility, which seems more likely, is that, while the widths of the peaks of Fig. \ref{peaks} are indeed related to the real parts of $s_{a}$ on general grounds, they are not exactly equal. On the other hand, we point out that, since the $s_{a}$ are related to the widths of the peaks, the transition in the scaling of the $\Delta_{k}$s at $\sigma=0$ suggests that the scaling of the heat-flux between the short-range and the long-range behaviour has to occur at $\sigma=0$.

\section{Conclusions} \label{sec6}

Heat transport in short-range linear systems has been widely studied \cite{Lepri_review}. On the contrary, the behaviour of linear oscillators with long-range power-law couplings is not yet well understood beyond the mean-field (fully-coupled) case \cite{defaveri2021,ucci2022}. In this paper, we have made a step forward along this direction by applying three different methods \cite{RLL, DHARREV, freitas2014analytic} that allow to compute numerically both the heat flux and its scaling with the system's size. All the methods give a clear scaling of the current with a power-law in the system's size. This scaling interpolates between the short-range behaviour, where the current is constant in the system's size, and the mean-field behaviour, where the current is inversely proporational to the system's size. However, the fitted scaling exponents show significant finite-size effects for all the three methods. The method of ref. \cite{RLL} which consists in solving a matricial equation is straightforwardly applicable to the long-range case. The Green's function approach allows to express the current as an integral over frequencies, which cannot be solved analytically. However, the integrand has the interesting property of showing a sequence of peaks that accumulate near the band edges of the spectrum. Further properties of these peaks can be inferred using the third method, which allows to compute the poles of the Green's function. Indeed, the real and the imaginary part of these poles are related to the position and the width of the peaks, respectively. We find a sharp transition in the scaling of the real parts of the poles at the value of the long-range coupling exponent $\sigma=0$ corresponding to the transition between the long-range and the short-range behaviour of the system. The crucial problem is now the dependence on $\sigma$ of the scaling exponent of the current. Assuming that all of the peaks of the integrand are well-separated Lorentzians and that their widths are exactly given by the real parts of the poles, we might conclude that the heat current scales as $\mathcal{J} \propto N^{-|\sigma|}$ for $-1<\sigma<0$.
c%compatible
in agreement with the one derived directly from the fit of the current,
which is anyway affected -- at least for small values of $|\sigma|$ --  
by significant finite-size effects. %This contradiction
The disagreement between these two scaling exponents remains to be explored,
%unexplained, however
even though our analysis of the scaling of the real part of the poles of the Green's
function %strongly
clearly supports the presence of a %sharp
transition at $\sigma=0$ from the long-range to the short-range behaviour.  \\

\begin{acknowledgments}
We gratefully thank Celia Anteneodo and Lucianno Defaveri for useful discussions.
SL and SR acknowledge partial support from project MIUR-PRIN2017 \textit{Coarse-grained description for non-equilibrium systems and transport phenomena (CO-NEST)} n. 201798CZL. 
\end{acknowledgments}

\bibliographystyle{apsrev4-1}

\end{document}